\documentclass[conference]{IEEEtran}
\ifCLASSINFOpdf
  \usepackage[pdftex]{graphicx}
\else
  \usepackage[dvips]{graphicx}
\fi
\usepackage[cmex10]{amsmath}
\usepackage{amssymb,dsfont}
\usepackage[tight,footnotesize]{subfigure}
\newcommand{\nr}{n}
\newcommand{\bfe}{\boldsymbol e}
\newcommand{\bfY}{\boldsymbol Y}
\newcommand{\bfomega}{\boldsymbol\omega}
\newcommand{\bfzero}{\boldsymbol 0}

\DeclareMathAlphabet{\mathbsf}{OT1}{cmss}{bx}{n}
\DeclareMathAlphabet{\mathssf}{OT1}{cmss}{m}{sl}
\DeclareMathOperator{\step}{step}
\DeclareMathOperator{\E}{E}
\DeclareMathOperator*{\argmax}{argmax}

\newcommand{\indi}[1]{\mathbb{I}_{\{#1\}}}

\begin{document}
%
\title{Balancing Lifetime and Classification Accuracy of Wireless Sensor Networks}

\author{\IEEEauthorblockN{Kush R. Varshney and Peter M. van de Ven}
\IEEEauthorblockA{Business Analytics and Mathematical Sciences Department\\
IBM Thomas J.~Watson Research Center\\
1101 Kitchawan Road, Route 134\\
Yorktown Heights, New York 10598\\
Email: \{krvarshn,pmvandev\}@us.ibm.com}}


%


\maketitle

\begin{abstract}
Wireless sensor networks are composed of distributed sensors that can be used for signal detection or classification. The likelihood functions of the hypotheses are often not known in advance, and decision rules have to be learned via supervised learning. A specific such algorithm is Fisher discriminant analysis (FDA), the classification accuracy of which has been previously studied in the context of wireless sensor networks. Previous work, however, does not take into account the communication protocol or battery lifetime of the sensor networks; in this paper we extend the existing studies by proposing a model that captures the relationship between battery lifetime and classification accuracy. In order to do so we combine the FDA with a model that captures the dynamics of the Carrier-Sense Multiple-Access (CSMA) algorithm, the random-access algorithm used to regulate communications in sensor networks. This allows us to study the interaction between the classification accuracy, battery lifetime and effort put towards learning, as well as the impact of the back-off rates of CSMA on the accuracy. We characterize the tradeoff between the length of the training stage and accuracy, and show that accuracy is non-monotone in the back-off rate due to changes in the training sample size and overfitting.

\end{abstract}


%
\IEEEpeerreviewmaketitle

\section{Introduction}
\label{sec:intro}

Wireless sensor networks are used for detection or classification, whether for surveillance, environmental monitoring, or any of the myriad other application domains that are emerging in the age of big data.  In many such applications, the likelihood functions of the hypotheses, e.g.,~the presence or absence of a particular physical phenomenon, are not known before the sensor network is deployed; in these applications, the sensor network requires training prior to operation via supervised learning \cite{NguyenWJ2005,PreddKP2006,PreddKP2006a,ZhengKP2011,VarshneyW2011}. The resulting classification accuracy improves with the number of measurements taken during training ~\cite{Varshney2012}, but increasing length of the training stage further reduces the limited battery capacity for the operational stage. Therefore, the amount of resources expended during training mediates operational lifetime and accuracy of the sensor network.

The energy consumption of sensor nodes, and thus the lifetime of the network, is dominated by energy expended on communication. Node transmissions in wireless sensor networks are commonly regulated by the Carrier-Sense Multiple-Access (CSMA) algorithm 
~\cite{AkyildizSSC2002,TayJB2004,HongV2007,SinghB2012}. This algorithm is implemented in TinyOS, a popular open source operating system for wireless sensor networks, and is part of the IEEE 802.15.4 standard for wireless sensor network communication \cite{802154}. Nodes using CSMA access the medium in a distributed manner, and wait some random back-off time between successive transmissions.

In this paper we consider a scenario where a set of measurements and classification is required every time unit. Only nodes that are active at that time perform a measurement and transmit the result, so the number of measurements collected varies over time. We develop and analyze a model of sensor networks that perform supervised classification \emph{in situ}, using the Fisher discriminant analysis (FDA) learning algorithm, with a training stage and an operational stage enabled by CSMA.  The specific analysis of focus is the relationship between operational accuracy and lifetime, which we show to be of a fundamentally different character than for the case of detection with known likelihood functions, due to overfitting.
In characterizing operational classification accuracy (in contrast to classification accuracy on training samples), we make use of generalization approximations for FDA developed by Raudys et al.~\cite{RaudysY2004}.

Battery capacity is characterized by the number of transmissions (and thus measurements) that can be performed, whether they be during training or operation. As every measurement corresponds to one transmission, the expected network lifetime is inversely proportional to the node throughput in our model.  The performance measures of interest are the classification accuracy and operational lifetime, which is the lifetime spent in the operational stage, not in the training stage.  The two main parameters available for configuring the sensor network are the CSMA back-off rates (the reciprocal of the mean back-off time), and the fraction of the lifetime spent in the training stage.

As the back-off rates of the nodes increase, states with many actively transmitting nodes are more likely. This requires more energy consumption, and also affect classification accuracy.  Classification accuracy is not monotonically increasing in the number of active nodes due to the phenomenon of overfitting, as we discuss in \cite{Varshney2012}.  We also show that operational accuracy as a function of back-off rate exhibits the hallmarks of overfitting in one regime, but in another regime, has a behavior quite different than any behavior usually encountered in statistical learning \cite{Vapnik1995}.

The analysis of supervised classification for sensor networks in the researcg literature is limited \cite{PreddKP2006}: Investigations have been predominantly concerned with the detection case where the likelihood functions are known.  Moreover, sensor network research tends to separate learning issues from the communication aspect.
There are several works that model CSMA communication in sensor networks generally, e.g.~\cite{FeoD2011} and references therein, but not with the supervised classification application as part of the formulation.  Cross-layer work that does consider the networking issues together with a detection or estimation application, e.g.,~the correlation-based collaborative MAC protocol \cite{VuranA2006}, is again focused on the case with known likelihoods.  So although FDA and the performance of CSMA-like algorithms has been widely studied in the research literature, we are the first to jointly consider classification accuracy and communication aspects of wireless sensor networks. 

We consider both the case of statistically independent and identically distributed (i.i.d.) measurements from different nodes, and the case of measurements exhibiting correlation that depends on the spatial distance between the nodes.  Having i.i.d.\ measurements is a common simplifying assumption in wireless sensor network detection \cite{ChamberlandV2003}.  A model with spatially-correlated measurements is much closer to reality in most applications \cite{VuranAA2004,JindalP2006}.  We assume that the learning algorithm has no prior information on the distribution and correlation of the measurements; 
the FDA has to estimate means and covariances as part of the training stage.  The spatial correlation is encoded via a Gauss--Markov random field (GMRF) model.

The CSMA model under consideration was first introduced in the 1980s in the context of packet radio networks~\cite{BK80,BKMS87} and was later applied to networks based on the IEEE 802.11 standard~\cite{WK05,LKLW08,DDT09,VandevenLDJ2012}. More recently, it has been used to study so-called adaptive CSMA algorithms, where the back-off rate of the nodes changes with their congestion level~\cite{RSS09a,BF10,GS10,JW10a}. Although the representation of binary exponential back-off mechanism in the above-mentioned models is far less detailed than in the landmark work of Bianchi~\cite{Bianchi00} and similar results focusing on sensor networks, e.g.,~\cite{FeoD2011,PKCCK05}, the general interference graph offers greater versatility and covers a broad range of topologies.

The remainder of the paper is organized as follows.  In Section~\ref{sec:fda}, we describe the setup of the sensor network system from the FDA supervised classification perspective and in Section~\ref{sec:csma}, we describe the setup of the sensor network system from the CSMA communication perspective.  In Section~\ref{sec:relation} we derive the relationship between operational lifetime and accuracy, and Section~\ref{sec:example} presents numerical results of lifetime and accuracy for two special cases, illustrating the complicated balancing act that is involved.  Section~\ref{sec:discuss} provides a discussion and several ideas for future directions of research.

\section{Fisher Discriminant Analysis}
\label{sec:fda}

Consider a sensor network consisting of $n$ sensor nodes each taking a scalar measurement combined into a joint measurement vector $\mathbf{x}_j \in \mathbb{R}^n$. In the general supervised classification problem, we are given $m$ sample pairs $\{(\mathbf{x}_1,y_1),\ldots,(\mathbf{x}_m,y_m)\}$ known as the training set, with measurement $\mathbf{x}_j$ and the class label or hypothesis $y_j \in \{0,1\}$.    The training samples are acquired by the network after deployment and before the operational stage.  The availability of labels for the training measurements is an assumption made in \cite{NguyenWJ2005,PreddKP2006,PreddKP2006a,ZhengKP2011} as well.  Once the training set is acquired, the samples are used to learn a classification function or decision rule $\hat{y}(\cdot)$ that will accurately classify new unseen and unlabeled samples $\mathbf{x}$ from the same distribution from which the training set was drawn.

In this paper, we focus on a simple, classical decision rule $\hat{y}$, the Fisher discriminant analysis classifier \cite{Fisher1936,Anderson1958}:
\begin{equation}
\label{eq:LDAstep}
	\hat{y}(\mathbf{x}) = \step(\mathbf{w}^T\mathbf{x} + \theta),
\end{equation}
where
\begin{align}
\nonumber	\mathbf{w} &= \left(\hat{\boldsymbol{\Sigma}}_0 + \hat{\boldsymbol{\Sigma}}_1 \right)^{-1}\left(\hat{\boldsymbol{\mu}}_1 - \hat{\boldsymbol{\mu}}_0\right), \\
	\theta &= -\tfrac{1}{2}\mathbf{w}^{T}\left(\hat{\boldsymbol{\mu}}_0 + \hat{\boldsymbol{\mu}}_1\right),
\end{align}
and $\hat{\boldsymbol{\mu}}_0$, $\hat{\boldsymbol{\mu}}_1$, $\hat{\boldsymbol{\Sigma}}_0$, and $\hat{\boldsymbol{\Sigma}}_1$ are the conditional sample means and covariances of the $m$ training samples.  The Fisher discriminant analysis rule is a plug-in classifier that follows from the likelihood ratio test for optimal signal detection between Gaussian signals with the same covariance and different means.  The rule \eqref{eq:LDAstep} is applied in the operational stage of the sensor network to classify new observations.

Given the FDA decision rule \eqref{eq:LDAstep}, we would like to characterize its performance, specifically its classification accuracy as it generalizes to new unseen samples in the operational stage.  Generalization accuracy, however, is a functional of the underlying data distribution $f_{\mathbsf{x},\mathssf{y}}(\mathbf{x},y)$, and we must first specify a probability distribution of the sensor measurements.  We employ the same GMRF statistical model for sensor measurements as \cite{Varshney2012,AnandkumarTS2009}. That is, the $n$ sensor nodes are deployed on the plane with spatial locations $\mathbf{v}_i \in \mathbb{R}^2$, $i= 1,\ldots,n$.  The likelihoods of the two hypotheses are Gaussian: $f_{\mathbsf{x}|\mathssf{y}}(\mathbf{x}|\mathssf{y} = 0) \sim \mathcal{N}(\boldsymbol{\mu}_0,\boldsymbol{\Sigma})$ and $f_{\mathbsf{x}|\mathssf{y}}(\mathbf{x}|\mathssf{y} = 1) \sim \mathcal{N}(\boldsymbol{\mu}_1,\boldsymbol{\Sigma})$.  The prior probabilities of the hypotheses are equal: $\Pr[\mathssf{y} = 0] = \Pr[\mathssf{y} = 1] = 1/2$.  For simplicity of exposition $\boldsymbol{\mu}_0 = \boldsymbol{0}$ (the vector of all zeroes) and $\boldsymbol{\mu}_1 = \boldsymbol{1}$ (the vector of all ones).

The covariance structure is based on the Euclidean nearest neighbor graph of the sensors: The (undirected) nearest neighbor graph contains an edge between sensor $i$ and sensor $i'$ if sensor $i$ is the nearest neighbor of sensor $i'$ or if sensor $i'$ is the nearest neighbor of sensor $i$.  The set of edges in the nearest neighbor graph is denoted $\mathcal{E}$.  The diagonal elements of $\boldsymbol{\Sigma}$ are all equal to $\sigma^2$.  The elements of $\boldsymbol{\Sigma}$ corresponding to edges in the nearest neighbor graph are:
\begin{equation}
\label{eq:correlationdecay}
	\{\boldsymbol{\Sigma}\}_{ii'} = \sigma^2g(d(\mathbf{v}_i,\mathbf{v}_{i'})), \quad (i,i') \in \mathcal{E},
\end{equation}
where $g(\cdot): \mathbb{R}^+ \rightarrow (0,1)$ is a decreasing function that encodes correlation decay with distance.  The inverse covariance matrix $\mathbf{J} = \boldsymbol{\Sigma}^{-1}$ is used to specify the remaining elements.  The off-diagonal elements of $\mathbf{J}$ corresponding to sensor pairs $(i,i')$ that do not have an edge in the nearest neighbor graph are zero, i.e.
\begin{equation}
\label{eq:Jzeros}
	\{\mathbf{J}\}_{ii'} = 0, \quad i \neq i',\, (i,i') \not\in \mathcal{E}.
\end{equation}
We also consider the case of i.i.d.~observations in the paper, in which case $g(d) = 1$ for $d=0$ and $g(d) = 0$ otherwise.

A highly accurate approximation of generalization accuracy $A = \Pr[\hat{y}(\mathbsf{x}) = \mathssf{y}]$ for the FDA decision rule as described above is found in \cite{Varshney2012}. Based on \cite{RaudysY2004}, this approximation is given as
\begin{equation}
\label{eq:raudys}
	A \approx \Phi\left(\frac{\delta}{2}\left[\left(1 + \frac{4n}{m\delta^2}\right)\frac{m}{m-n}\right]^{-\tfrac{1}{2}}\right), \quad m > n,
\end{equation}
where $\Phi(\cdot)$ is the Gaussian cumulative distribution function and $\delta$ is known as the Mahalanobis distance
\begin{equation}
\label{eq:mahalanobis1a}
	\delta^2 = \frac{n}{\sigma^2} - \frac{2}{\sigma^2}\sum_{(i,i')\in\mathcal{E}} \frac{g(d(\mathbf{v}_i,\mathbf{v}_{i'}))}{1 + g(d(\mathbf{v}_i,\mathbf{v}_{i'}))}.
\end{equation}
In case $m \le n$ there are insufficient training samples for accurate classification and we have $A = 0.5$. In the i.i.d.~case, $\delta^2$ simplifies to $\frac{n}{\sigma^2}$.

\section{Carrier-Sense Multiple-Access}
\label{sec:csma}

The CSMA algorithm is an example of a random-access algorithm, where nodes decide for themselves when to transmit, based on local information only. We assume that the $\nr$ nodes share the wireless medium according to a CSMA-type protocol.

The network is described by an undirected conflict graph $(\mathcal{V}, \mathcal{E})$, where the set of vertices $\mathcal{V} = \{1, \dots, \nr\}$ represents the nodes of the network and the set of edges $\mathcal{E} \subseteq \mathcal{V} \times \mathcal{V}$ indicates which pairs of nodes cannot activate simultaneously. For ease of presentation we assume that the conflict graph is the same as the nearest neighbor graph introduced in Section~\ref{sec:fda}. Nodes that are neighbors in the conflict graph are prevented from simultaneous activity by the carrier-sensing mechanism. An inactive node is said to be blocked whenever any of its neighbors is active, and unblocked otherwise.

The transmission times of node~$i$ are independent and exponentially distributed with unit mean. When node~$i$ is blocked it remains silent until all its neighbors are inactive, at which point it tries to activate after an exponentially distributed back-off time with mean $1/\nu_i$.

The set $\Omega$ of all feasible joint activity states of the network in this case corresponds to the incidence vectors of all independent sets of the conflict graph. Let the network state at time~$t$ be denoted by $\bfY(t) = (Y_1(t),Y_2(t),\dots,Y_{\nr}(t)) \in \Omega$, with $Y_i(t)$ indicating whether node~$i$ is active at time~$t$ ($Y_i(t) = 1$) or not ($Y_i(t) = 0$). Then $\{\bfY(t)\}_{t \ge 0}$ is a Markov process which is fully specified by the state space $\Omega$ and the transition rates
\begin{equation}
r(\omega,\omega')=\left\{
                    \begin{array}{ll}
                      \nu_i, & {\rm if~} \bfomega'=\bfomega + \bfe_i \in \Omega, \\
                      1, & {\rm if~} \bfomega'=\bfomega - \bfe_i \in \Omega, \\
                      0, & {\rm otherwise}.
                    \end{array}
                  \right.
\end{equation}
Here $\bfe_i$ denotes the vector of length $\nr$ with all zeroes except for a 1 at position $i$.

Since $\bfY(t)$ is reversible (see~\cite{BK80}), the following product-form stationary distribution $\pi$ exists:
\begin{equation}\label{eqn:lim_dist}
\pi(\bfomega) =  \left\{
         \begin{array}{ll}
           Z^{-1}\prod_{i=1}^{\nr}\nu_i^{\omega_i}, & \hbox{if~} \bfomega \in \Omega, \\
           0, & \hbox{otherwise,}
         \end{array}
       \right.
\end{equation}
where
\begin{equation}\label{eqn:partition_function_general}
Z = \sum_{\bfomega \in \Omega} \prod_{i = 1}^\nr \nu_i^{\omega_i}
\end{equation}
is the normalization constant that makes $\pi$ a probability measure.

The rate $\theta_i$ at which sensor node~$i$ makes observations (or, alternatively, the rate at which it does transmissions) is referred to as the throughput of this node, and may be written as
\begin{equation}\label{eqn:throughput}
\theta_i = \sum_{\bfomega \in \Omega} \pi(\bfomega) \indi{\omega_i = 1}.
\end{equation}
Sensor nodes rely on batteries for energy, and we assume that all nodes have a battery that allows them to make $l$ transmissions each before their battery is drained. Consequently, the expected lifetime of a node can be written as
\begin{equation}
\label{eq:lifetime}
	T_i = \frac{l}{\theta_i}.
\end{equation}
The activity process in the training stage is the same as in the operational stage.  We denote by $0\le \alpha \le 1$ the fraction of the battery capacity that is dedicated to training the sensor network. So the testing lifetime of node~$i$ is $\alpha T_i$, and the operational lifetime (the quantity we would like to be large) is:
\begin{equation}
\label{eqn:operlifetime}
	U_i = (1-\alpha)T_i = (1-\alpha)\frac{l}{\theta_i}.
\end{equation}

The model we have specified is fully general for any $n$-node conflict graph.  We work with this general model throughout the remainder of the paper, but also focus on two illustrative special cases.  The two special cases of the CSMA network we consider are an $n$-node network where all networks are disjoint and a three-node linear network.

\subsection{Independent Nodes}

First, consider an $n$-node network where all nodes can be active simultaneously. This corresponds to an interference graph with an empty edge set $\E = \phi$. We have $\Omega = \{0,1\}^n$ and set $\nu_i \equiv \nu$ so the stationary distribution~\eqref{eqn:lim_dist} simplifies to
\begin{equation}
\label{eq:stationary1}
	\pi(\boldsymbol{\omega}) = \frac{1}{(\nu+1)^n}\nu^{\|\boldsymbol{\omega}\|_1}.
\end{equation}
The stationary probability of any particular state only depends on the number of active nodes in that state and on the back-off rate $\nu$.  Thus, for notational convenience, we introduce $\pi(k)$ as the stationary probability of being in any state with $k$ active nodes, and we write
\begin{equation}
\label{eq:stationary2}
	\pi(k) = \frac{1}{(\nu+1)^n}\binom{n}{k}\nu^{k},
\end{equation}
which follows since there are $\binom{n}{k}$ different activity states with $k$ nodes transmitting.

With equal back-off rates and disjoint nodes, the stationary throughput~\eqref{eqn:throughput} is the same for all nodes
\begin{equation}
\label{eq:throughput1}
	\theta_i \equiv \theta = \frac{\nu}{\nu+1}.
\end{equation}
Moreover, all nodes have the same lifetime, and the operational lifetime of the network may be written as
\begin{equation}
\label{eq:lifetime1}
	U_i\equiv U = (1-\alpha)l \frac{\nu+1}{\nu}.
\end{equation}

\subsection{A Three-Node Linear Network}\label{sec:three-node}

Consider the three-node network where the nodes are positioned such that the carrier-sensing mechanism prevents node~2 from activating while either node 1 or node 3 is active. Nodes 1 and 3 can be active simultaneously, but their observations are correlated. The network can take five possible states
\begin{equation}
\Omega = \{\bfzero,\bfe_1,\bfe_2,\bfe_3,\bfe_1 + \bfe_3\}.
\end{equation}
Using~\eqref{eqn:lim_dist} we compute the following stationary probabilities:
\begin{align}
\nonumber \pi(\bfzero) &= Z^{-1},\\
\nonumber \pi(\bfe_i) &= Z^{-1}\nu_i, \quad i = 1,2,3,\\
\pi(\bfe_1 + \bfe_3) &= Z^{-1}\nu_1 \nu_3.
\end{align}

In order to make sure that all nodes have the same throughput and  lifetime, we fix some parameter $\eta > 0$ and choose $\nu_1 = \nu_3 = \eta$ and $\nu_2 = \eta (\eta + 1)$. So node~2 has a shorter mean back-off time in order to compensate for its disadvantageous position in the network, and all nodes have throughput (see~\cite{VandevenLDJ2012})
\begin{equation}
\theta_i \equiv \theta = \frac{\eta}{2 \eta + 1}
\end{equation}
and operational lifetime
\begin{equation}
U_i \equiv U = (1-\alpha)l \frac{2 \eta + 1}{\eta}.
\end{equation}
The normalization constant with these back-off rates is given by
\begin{equation}
	Z = 2\eta^2 + 3\eta + 1.
\end{equation}

\section{Relationship Between Lifetime and Accuracy}
\label{sec:relation}

We are now in position to combine the FDA model from Section~\ref{sec:fda} and the CSMA model presented in Section~\ref{sec:csma} to derive the relationship between generalization accuracy and operational lifetime. This is mediated by two parameters: the back-off rate $\nu$ or $\eta$ and the fraction of the lifetime spent in the training stage $\alpha$.

Due to the interference constraints and the intermittent nature of CSMA communications, not all nodes produce and validly communicate measurements at all times. So the training samples are acquired under different activity states $\bfomega \in \Omega$. Thus studying the relationship between accuracy and lifetime is not simply a matter of joining the corresponding expressions~\eqref{eq:raudys} and~\eqref{eqn:operlifetime}.

This issue of incomplete data due to the activity process can be addressed in several ways, including data imputation \cite{GelmanH2007}.  Although various elaborate schemes are available, they come at the cost of additional computation, communication, and coordination that are at a premium in the sensor network setting. Instead, we choose to model the classification by having separately learned classifiers for different activity states. In the operational stage the appropriate classifier is used for prediction based on the activity state of the measurements. In this setup, we associate with each state $\bfomega$ a number of training samples
\begin{equation}
\label{eq:msubomega}
	m_{\boldsymbol{\omega}} = \alpha T \pi(\boldsymbol{\omega}).
\end{equation}
Then we compute the overall generalization accuracy as the weighted sum of the individual generalization accuracies for each pattern according to their stationary probabilities:
\begin{equation}
\label{eq:generror1}
	A \approx \sum_{\boldsymbol{\omega}\in\Omega} \pi(\boldsymbol{\omega}) \Phi\left(\frac{\delta}{2}\left[\left(1 + \frac{4\|\boldsymbol{\omega}\|_1}{m_{\boldsymbol{\omega}}\delta^2}\right)\frac{m_{\boldsymbol{\omega}}}{m_{\boldsymbol{\omega}}-\|\boldsymbol{\omega}\|_1}\right]^{-\tfrac{1}{2}}\right),
\end{equation}
with $\pi$ the stationary distribution~\eqref{eqn:lim_dist} and $m_{\boldsymbol{\omega}}$ as in~\eqref{eq:msubomega}.

We now compute the generalization accuracies for the two special cases introduced in Section~\ref{sec:csma} with the GMRF of the measurements having the same graph structure as the CSMA network.

\subsection{Independent Nodes}
\label{sec:relation:iid}

As discussed in Section~\ref{sec:csma} for a set of $n$ disjoint nodes, all patterns with $k$ active nodes have the same stationary probability $\pi(k)$ given in \eqref{eq:stationary2}, and all nodes have equal throughput~\eqref{eq:throughput1} and lifetime~\eqref{eq:lifetime1}. We denote by $m_k$ the number of training samples for patterns with $k$ active nodes, and by summing~\eqref{eq:msubomega} over all states with $k$ active nodes, we write
\begin{equation}
\label{eq:mdisjoint}
	m_k	= \alpha l \binom{n}{k} \frac{\nu^{k-1}}{(\nu+1)^{n-1}}.
\end{equation}
As discussed in Section~\ref{sec:fda}, with i.i.d.~measurements from $n$ sensors, the squared Mahalanobis distance is $\frac{n}{\sigma^2}$.  Thus, with $k$ active sensors, the squared Mahalanobis distance is $\frac{k}{\sigma^2}$.  Substituting the expression for the stationary distribution~\eqref{eq:stationary2} and the number of training samples~\eqref{eq:mdisjoint} into the expression for the generalization accuracy~\eqref{eq:generror1} we obtain
\begin{multline}
\label{eq:generror2}
	A \approx \\ \frac{1}{(\nu+1)^n}\sum_{k=0}^{n} \binom{n}{k}\nu^k \Phi\left(\frac{\sqrt{k}}{2\sigma}\left[\left(1 + \frac{4\sigma^2}{m_k}\right)\frac{m_k}{m_k-k}\right]^{-\tfrac{1}{2}}\right).
\end{multline}

\subsection{A Three-Node Linear Network}
\label{sec:relation:3node}

Recall from Section~\ref{sec:three-node} that the three-node network has 5 feasible states. The four non-empty states have squared Mahalanobis distance
\begin{align}
\nonumber	\delta^2_{\bfe_i}  &= \frac{1}{\sigma^2}, \quad i = 1,2,3,\\
 \delta^2_{\bfe_1 + \bfe_3} &= \frac{2}{\sigma^2}\cdot\frac{1}{1 + g(d(\mathbf{v}_1,\mathbf{v}_2))g(d(\mathbf{v}_2,\mathbf{v}_3))}.
\end{align}
Note that since $g < 1$, the Mahalanobis distance of the larger state is larger than that of the states with only one node active, and is more valuable.

Evaluating~\eqref{eq:msubomega} we obtain an expression for the number of training samples for each state:
\begin{align}
\label{eqn:threenodem}
\nonumber 	m_{\bfzero} &= \alpha l \frac{1}{\eta^2+\eta}\\
\nonumber 	m_{\bfe_i} &= \alpha l \frac{1}{\eta+1}, \quad i = 1,3,\\
\nonumber    m_{\bfe_2} &= \alpha l, \\
     m_{\bfe_1 + \bfe_3} &= \alpha l \frac{\eta}{\eta+1}.
\end{align}

By weighting the individual generalization accuracies~\eqref{eq:generror1}, we obtain
\begin{multline}
\label{eqn:threenodeA}
	A \approx \frac{1}{2\eta^2 + 3\eta + 1}\Bigg[\frac{1}{2} + \\
\eta\Phi\left(\frac{1}{2\sigma}\left[\left(1 + \frac{4\sigma^2}{m_{\bfe_1}}\right)\frac{m_{\bfe_1}}{m_{\bfe_1}-1}\right]^{-\tfrac{1}{2}}\right) + \\
(\eta^2 + \eta)\Phi\left(\frac{1}{2\sigma}\left[\left(1 + \frac{4\sigma^2}{m_{\bfe_2}}\right)\frac{m_{\bfe_2}}{m_{\bfe_2}-1}\right]^{-\tfrac{1}{2}}\right) + \\
\eta\Phi\left(\frac{1}{2\sigma}\left[\left(1 + \frac{4\sigma^2}{m_{\bfe_3}}\right)\frac{m_{\bfe_3}}{m_{\bfe_3}-1}\right]^{-\tfrac{1}{2}}\right) + \\
\eta^2\Phi\left(\frac{\delta_{\bfe_1 + \bfe_3}}{2}\left[\left(1 + \frac{8}{m_{\bfe_1 + \bfe_3}\delta^2_{\bfe_1 + \bfe_3}}\right)\frac{m_{\bfe_1 + \bfe_3}}{m_{\bfe_1 + \bfe_3}-2}\right]^{-\tfrac{1}{2}}\right)\Bigg].
\end{multline}

\section{Examples}
\label{sec:example}

In Section~\ref{sec:relation} we derived the operational lifetime $U$, the number of training samples $m_{\bfomega}$ and the operational classification accuracy $A$ for a wireless sensor network with random-access communication as a function of the back-off rate $\nu$ and the fraction of the lifetime spent in training $\alpha$. Here we numerically evaluate these quantities for the special cases of independent nodes and the three-node linear network.  We include a comparison to the Bayes optimal detector with known likelihood functions and see that the accuracy behavior is markedly different.  Additionally, we see that there are two different regimes in the accuracy behavior as a function of the back-off rate, the second regime different than that usually seen in statistical learning.  The overall behavior is unique due to the combination of CSMA and FDA.

\subsection{Independent Nodes}
\label{sec:example:iid}

\begin{figure}
	\begin{center}
		\begin{tabular}{cc}
			\includegraphics[width=0.47\textwidth]{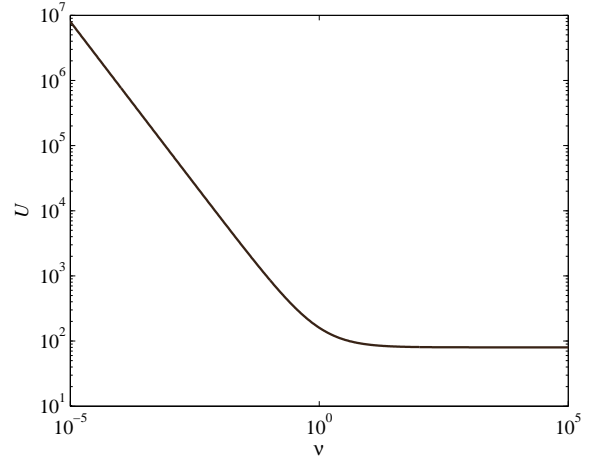}
		\end{tabular}
	\end{center}
	\caption{Operational lifetime as a function of back-off rate.}
	\label{fig:indep_Unu}
\end{figure}

\begin{figure}
	\begin{center}
		\begin{tabular}{cc}
			\includegraphics[width=0.47\textwidth]{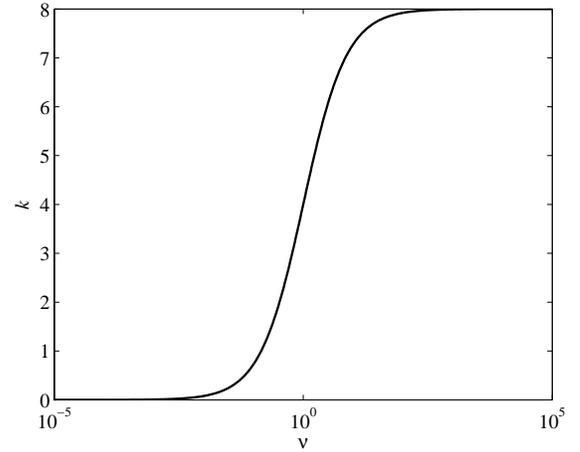}
		\end{tabular}
	\end{center}
	\caption{Expected number of active sensors as a function of back-off rate.}
	\label{fig:indep_knu}
\end{figure}

We consider a network of $n=8$ independent nodes with $l=100$ transmissions allowed by the battery per node.  The sensor measurement noise variance is set to $\sigma^2 = 1$.  Other parameter settings produce qualitatively similar results.  First, in Fig.~\ref{fig:indep_Unu}, we plot the operational lifetime $U$ as a function of the back-off rate $\nu$ for a fixed lifetime fraction devoted to training: $\alpha = 0.2$.
The operational lifetime is very high with low back-off rate because the system is mostly in the state with $k = 0$ active sensors, which does not drain sensor batteries at all.  Once states with more active sensors become more probable with increasing back-off rate, the lifetime drops rapidly to $U = (1 - \alpha)l$, the lifetime in the case where all nodes are always active. Fig.~\ref{fig:indep_knu} shows the expected number of active sensors $\bar{k}$ as a function of $\nu$.

\begin{figure}
	\begin{center}
		\begin{tabular}{cc}
			\includegraphics[width=0.47\textwidth]{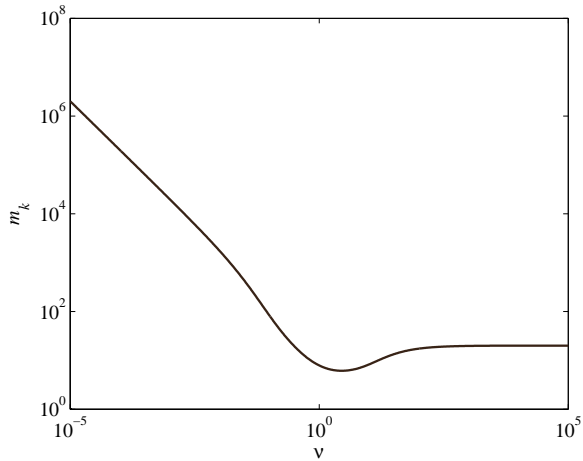}
		\end{tabular}
	\end{center}
	\caption{Expected number of training samples per state classifier as a function of back-off rate.}
	\label{fig:indep_mknu}
\end{figure}

One of the components of the generalization accuracy expression \eqref{eq:generror2} is the number of active sensors $k$; the other is $m_k$, the number of training samples. In Fig.~\ref{fig:indep_mknu} we set $\alpha = 0.2$, and plot $\bar{m}_k$, the weighted average over $k$ of $m_k$:
\begin{equation}
\bar{m}_k = \sum_{k = 0}^n \pi(k) m_k.
\end{equation}
Interestingly, this number is not monotonically decreasing as a function of $\nu$ like we see with the operational lifetime.  This is because when several different states all have non-negligible probability, the acquired training samples get divided to all of the different states.  Initially the number of training samples is very high because almost all of the training samples are for the state with no active sensors.  For large $\nu$ the number of training samples approaches $\alpha l$.

\begin{figure}
	\begin{center}
		\begin{tabular}{cc}
			\includegraphics[width=0.47\textwidth]{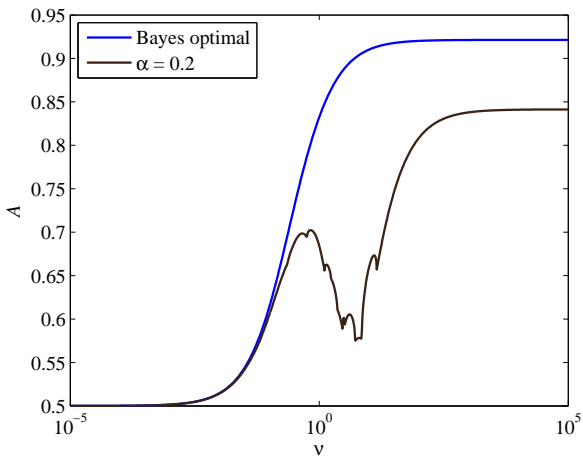}
		\end{tabular}
	\end{center}
	\caption{Operational classification accuracy as a function of back-off rate.}
	\label{fig:indep_Anu}
\end{figure}

\begin{figure}
	\begin{center}
		\begin{tabular}{cc}
			\includegraphics[width=0.47\textwidth]{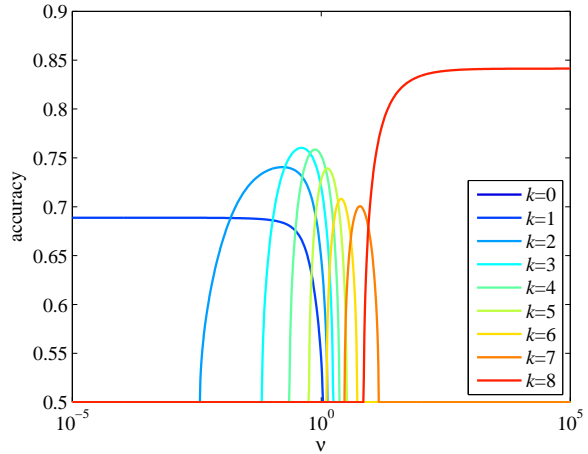}
		\end{tabular}
	\end{center}
	\caption{Operational classification accuracy of the different state classifiers as a function of back-off rate.}
	\label{fig:indep_Anustates}
\end{figure}

Now that we have looked at $\bar{k}$ and $\bar{m}_k$, we now examine the accuracy $A$ as a function of $\nu$, plotted in Fig.~\ref{fig:indep_Anu} for $\alpha = 0.2$.
The figure also shows the detection accuracy of the Bayes optimal decision rule with known likelihood functions.  The Bayes optimal accuracy is monotonically increasing in $\nu$, following the expected value of $k$.  On the other hand, the FDA classification accuracy first increases in $\nu$, starts decreasing with local bumps, and then increases.  The local bumps arise from the generalization accuracy behavior for different states, which are shown in Fig.~\ref{fig:indep_Anustates}.
Specifically, the figure shows the different $\Phi(\cdot)$ components of $A$ given in \eqref{eq:generror2}; these are functions of $\nu$ because the $m_k$ are.

The phenomenon of overfitting, demonstrated in \cite{Varshney2012}, is that for a fixed number of training samples and an increasing number of sensors, the generalization accuracy first increases and then decreases. Conversely, for a fixed number of sensors and an increasing number of training samples, the generalization accuracy monotonically increases.  With the wireless sensor network with CSMA communication, both of these effects intermingle as a function of $\nu$ because both the number of active sensors and the number of training samples changes.  The initial increase and decrease in $A$ is the manifestation of overfitting, where the generalization accuracy is best around $k = 3$ and $k = 4$.  For large $\nu$, the number of active sensors is essentially fixed at $k=n$ (seen in Fig.~\ref{fig:indep_knu}) and the number of training samples increases (seen in Fig.~\ref{fig:indep_mknu}), resulting in improving classification accuracy.

\begin{figure}
	\begin{center}
		\begin{tabular}{cc}
			\includegraphics[width=0.47\textwidth]{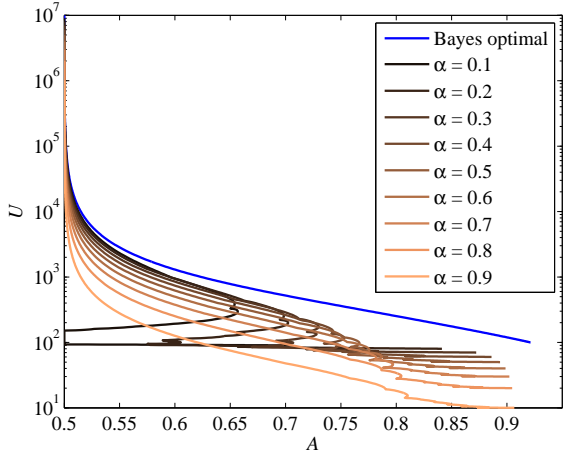}
		\end{tabular}
	\end{center}
	\caption{Relationship between operational lifetime and operational classification accuracy.}
	\label{fig:indep_UA}
\end{figure}

Finally, we examine the relationship between lifetime and accuracy in Fig.~\ref{fig:indep_UA}.
For comparison, the figure shows the relationship for the Bayes optimal decision rule, in which there is no lifetime devoted to training, only to operation. All curves represent parametric functions of $\nu$, and correspond to different values of $\alpha$ ranging from $0.1$ to $0.9$.  Different values of $\alpha$ contribute to the frontier of the relationship, the parts of the curve closest to the Bayes decision rule and closest to the top right corner of the plot.  At the extreme of random guessing, i.e.~$A = 0.5$, lifetime is maximized by not doing any training, i.e.~$\alpha=0$.  Small but increasing values of $\alpha$ then contribute to the frontier until a point when smaller values of $\alpha$ abruptly again become part of the frontier.  Very large values of $\alpha$ contribute to the frontier only when the very best accuracies possible are desired.

\subsection{A Three-Node Linear Network}
\label{sec:example:3node}

\begin{figure}
	\begin{center}
		\begin{tabular}{cc}
			\includegraphics[width=0.47\textwidth]{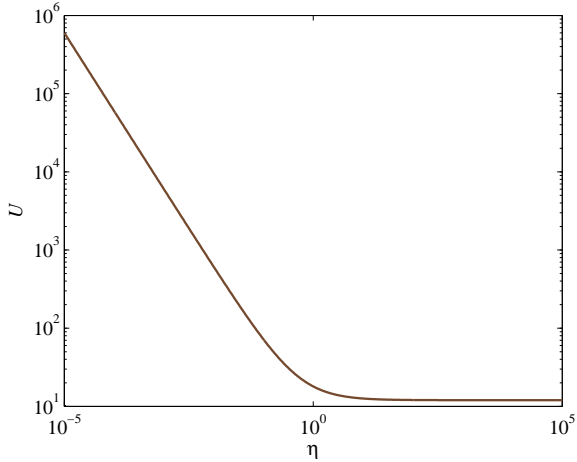}
		\end{tabular}
	\end{center}
	\caption{Operational lifetime as a function of back-off rate.}
	\label{fig:3node_Ueta}
\end{figure}

\begin{figure}
	\begin{center}
		\begin{tabular}{cc}
			\includegraphics[width=0.47\textwidth]{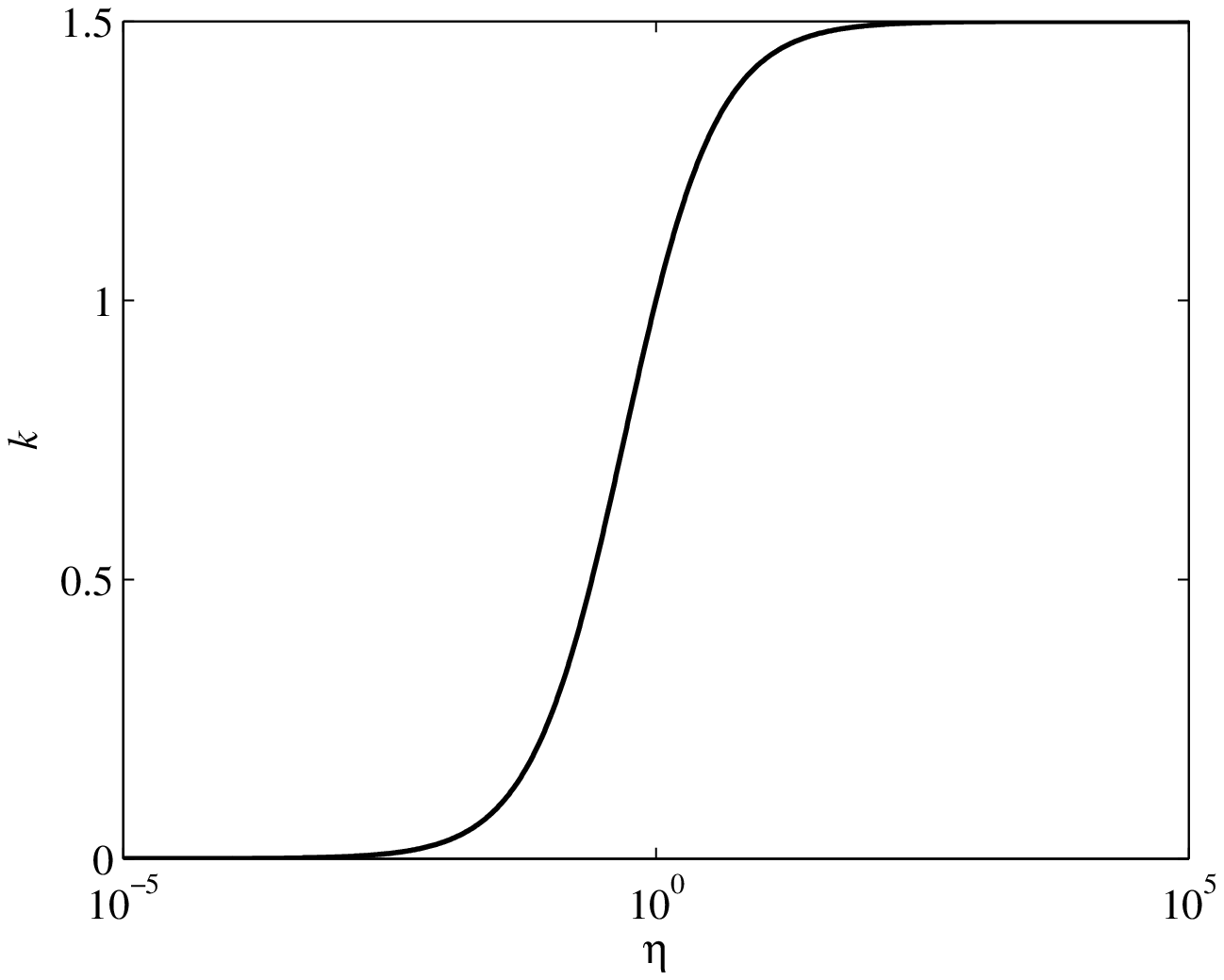}
		\end{tabular}
	\end{center}
	\caption{Expected number of active sensors as a function of back-off rate.}
	\label{fig:3node_keta}
\end{figure}

\begin{figure}
	\begin{center}
		\begin{tabular}{cc}
			\includegraphics[width=0.47\textwidth]{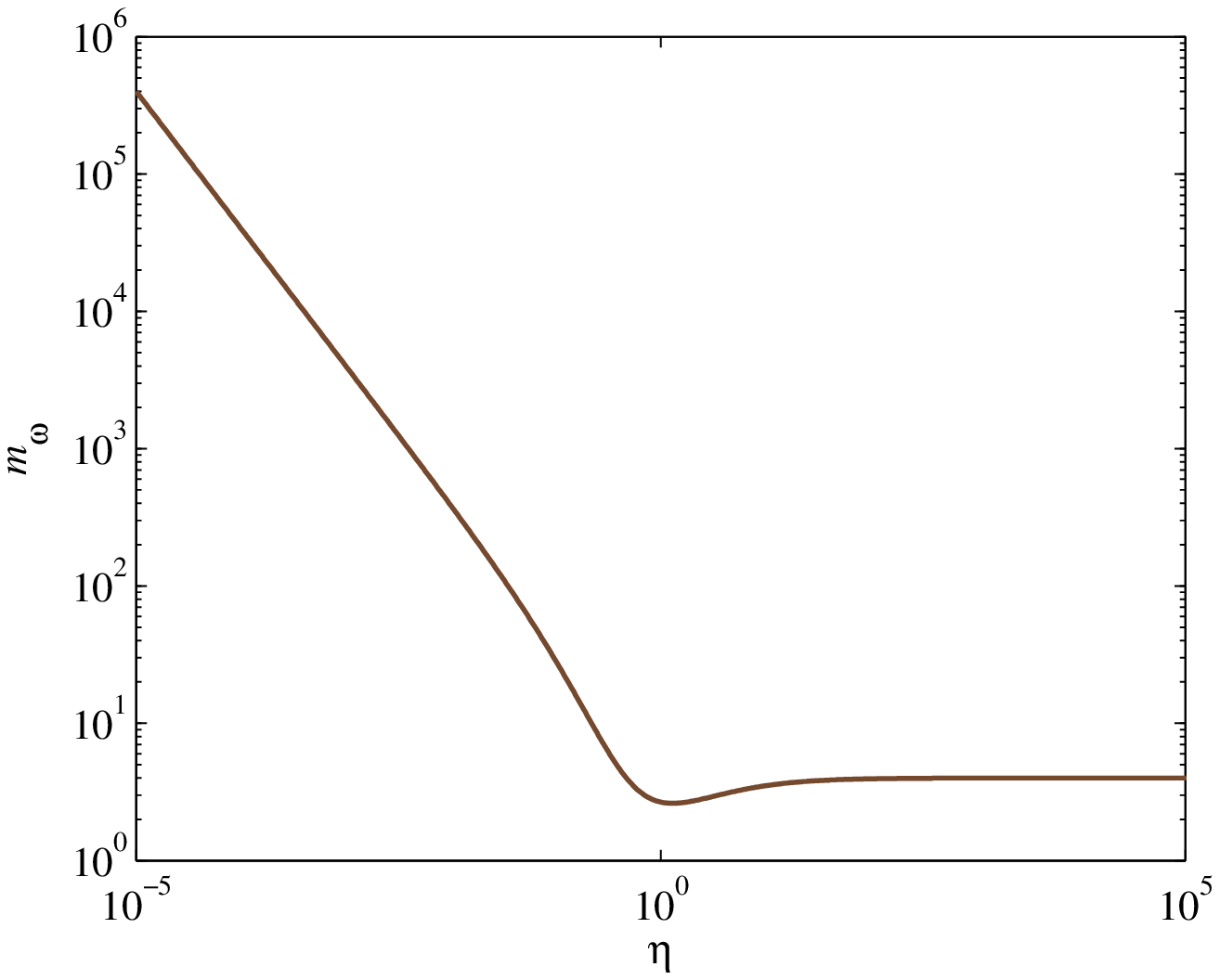}
		\end{tabular}
	\end{center}
	\caption{Expected number of training samples per state classifier as a function of back-off rate.}
	\label{fig:3node_momegaeta}
\end{figure}

\begin{figure}
	\begin{center}
		\begin{tabular}{cc}
			\includegraphics[width=0.47\textwidth]{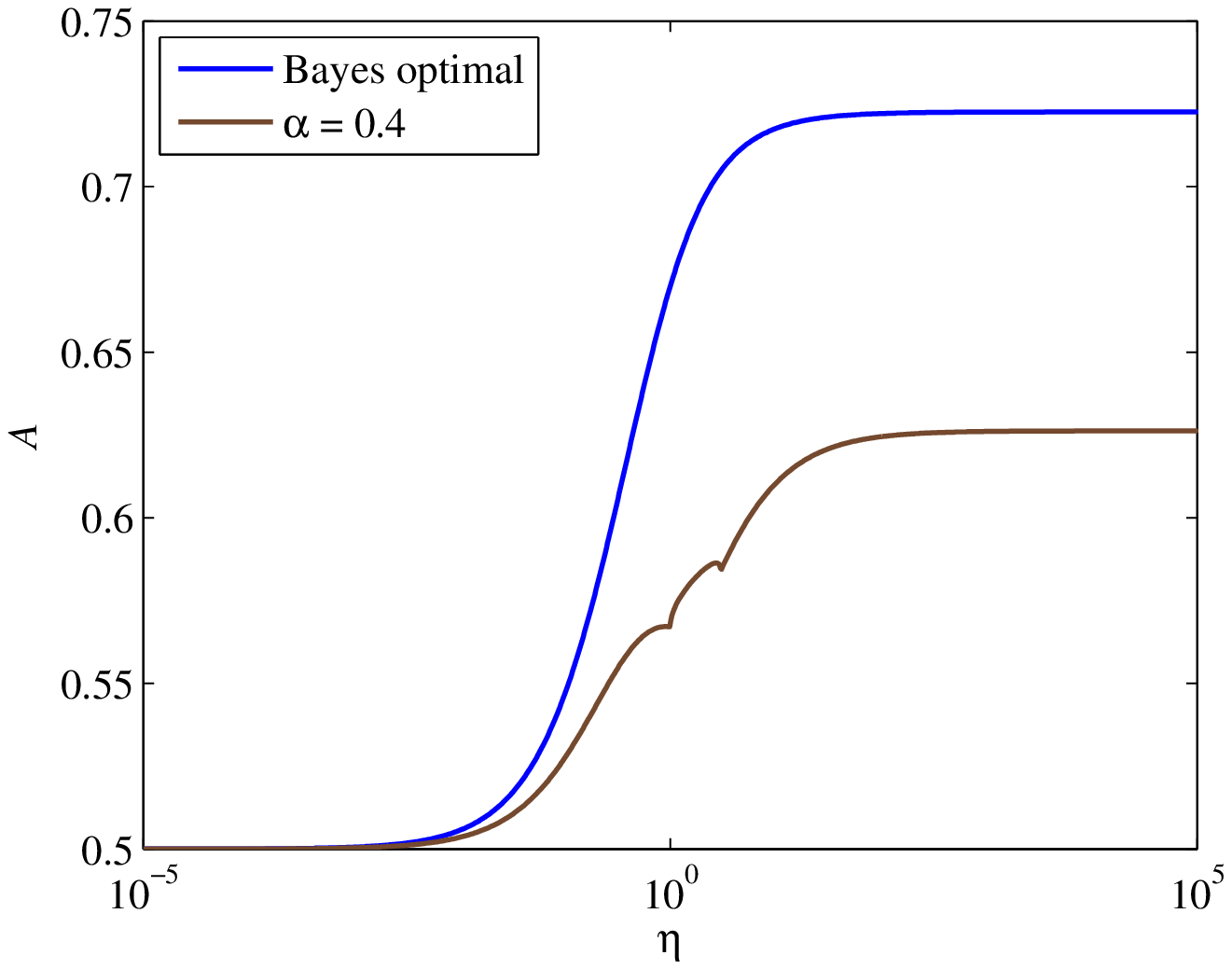}
		\end{tabular}
	\end{center}
	\caption{Operational classification accuracy as a function of back-off rate.}
	\label{fig:3node_Aeta}
\end{figure}

\begin{figure}
	\begin{center}
		\begin{tabular}{cc}
			\includegraphics[width=0.47\textwidth]{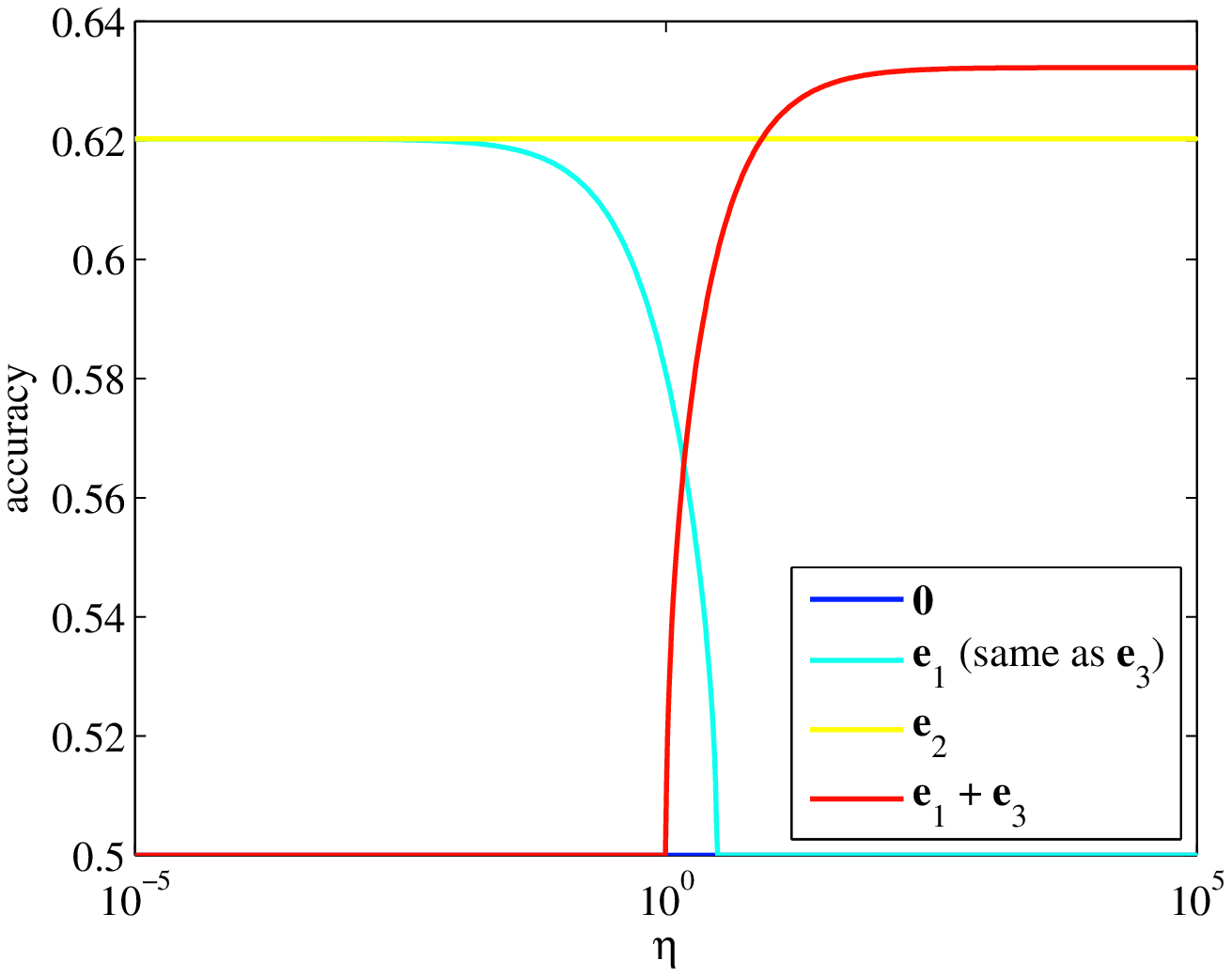}
		\end{tabular}
	\end{center}
	\caption{Operational classification accuracy of the different state classifiers as a function of back-off rate.}
	\label{fig:3node_Aetastates}
\end{figure}

\begin{figure}
	\begin{center}
		\begin{tabular}{cc}
			\includegraphics[width=0.47\textwidth]{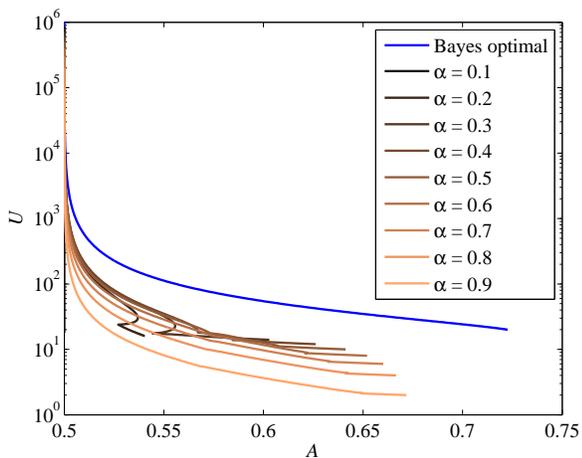}
		\end{tabular}
	\end{center}
	\caption{Relationship between operational lifetime and operational classification accuracy.}
	\label{fig:3node_UA}
\end{figure}

Having seen quite interesting behaviors for independent nodes, we now turn to a three-node linear network with correlated measurements and conflict graph preventing sensors 1 and 2, and sensors 2 and 3 from transmitting simultaneously.  We present similar plots as in Section~\ref{sec:example:iid}, with $l = 10$ and $\sigma^2 = 1$.  We present results for $g(d(\mathbf{v}_1,\mathbf{v}_2)) = g(d(\mathbf{v}_2,\mathbf{v}_3)) = \frac{1}{4}$ as the distance-based correlations.  For the dependent linear network case, we see more or less the same behavior as for the independent nodes in Fig.~\ref{fig:3node_Ueta}--Fig.~\ref{fig:3node_UA}.  Fig.~\ref{fig:3node_Ueta}, Fig.~\ref{fig:3node_momegaeta}, and Fig.~\ref{fig:3node_Aetastates} are given for $\alpha = 0.4$.  One difference from the independent nodes case is that for the $\bfe_2$ state, $m_{\bfomega}$ is constant and not a function of $\eta$.

\section{Conclusion and Outlook}
\label{sec:discuss}

In this paper we proposed a model to investigate the interaction between generalization accuracy and operational lifetime in wireless sensor networks. We demonstrate that this relationship is highly nontrivial, due to the joint effects of overfitting, the number of training samples and the changing weights of the various states. 
The two special cases for which we provide result plots are qualitatively similar, and changing the conflict graph and spatial correlation does not affect the general behavior.

For small increasing back-off rates, the accuracy improves until peaking.  At intermediate back-off rates, the accuracy gets worse due to overfitting, and then improves again for large back-off rates due to increasing training samples per state.  Due to these different regimes of increasing and decreasing accuracy in the back-off rate, along with different values of the fraction of lifetime to spend in training affecting both the operational lifetime and the accuracy, setting the parameters $\nu$ and $\alpha$ to achieve certain target performance is not straightforward.  The parameterized curves in Fig.~\ref{fig:indep_UA} and Fig.~\ref{fig:3node_UA} give (not necessarily intuitive) recommendations for balancing lifetime and accuracy as a function of the back-off rate and training fraction parameters.

The classification and communication models we have used, i.e.~FDA with GMRF-dependent sensor measurements and binary exponential back-off mechanism, are certainly simplified, but are general and amenable to analysis.  The guidelines and behaviors we see will transfer over in a broad sense to other classifiers and other similar random-access communication protocols.  The accuracy behavior that we see is partly due to the way we deal with measurements from different states through separate classifiers, but the general complicated behavior ought to remain if we take another approach.  



\subsection{Outlook}
\label{sec:conclusion}



\begin{figure}
	\begin{center}
		\begin{tabular}{cc}
			\includegraphics[width=0.47\textwidth]{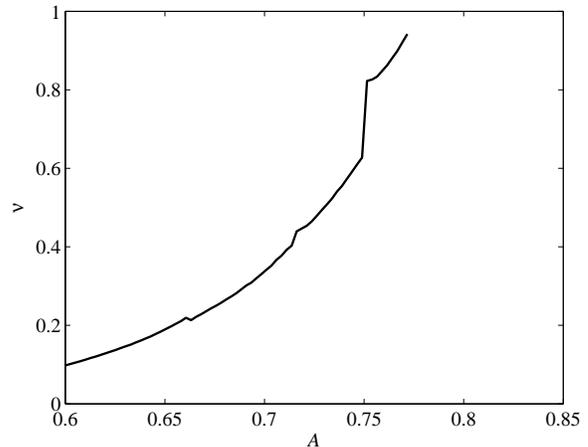}
		\end{tabular}
	\end{center}
	\caption{Plotting $\nu^*$ as a function of the desired accuracy.}
	\label{fig:allerton_indep_optnu}
\end{figure}
\begin{figure}
	\begin{center}
		\begin{tabular}{cc}
			\includegraphics[width=0.47\textwidth]{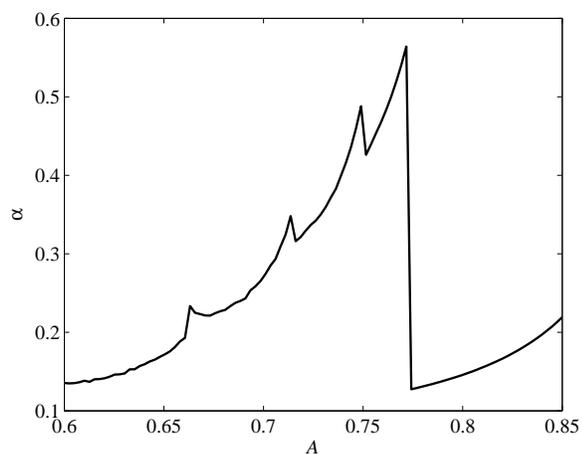}
		\end{tabular}
	\end{center}
	\caption{Plotting $\alpha^*$ as a function of the desired accuracy.}
	\label{fig:allerton_indep_optalpha}
\end{figure}

Having made the connection between lifetime and accuracy in Fig.~\ref{fig:indep_UA} and Fig.~\ref{fig:3node_UA}, the next step is to find the values of $\alpha$ and $\nu_i$ that achieve a certain target performance. For example, we may want to maximize the lifetime of the network, subject to certain accuracy constraints $\beta \in (0,1)$:
\begin{align}
\nonumber (\alpha^*,\nu^*) &= \argmax U(\alpha,\nu_1,\dots,\nu_n)\\
&{\rm s.t.~} A(\alpha,\nu_1,\dots,\nu_n) \ge \beta. \label{eqn:optimization}
\end{align}
The optimization problem~\eqref{eqn:optimization} is non-convex (as illustrated in Fig.~\ref{fig:indep_UA}), and we may approximate its solution using numerical methods. Some preliminary results are shown in Figs.~\ref{fig:allerton_indep_optnu} and~\ref{fig:allerton_indep_optalpha}, where plot the solution to~\eqref{eqn:optimization} for increasing $\beta$, in the model with $n$ independent nodes with the parameters as in Section~\ref{sec:example}. Fig.~\ref{fig:allerton_indep_optnu} shows that the $\nu$ increases almost monotonically, and jumps to infinity around $\beta = 0.772$. In practice we see that the back-off rate is constrained by physical limitations and by the communication protocol, so $\nu$ is bounded from above. Fig.~\ref{fig:allerton_indep_optalpha} shows a more irregular behavior for $\alpha$, with a sharp drop when $\nu$ jumps to infinity.


The non-monotonicity of the classification accuracy in the back-off rates makes an analytic approach to optimization difficult, and an alternative solution would be to approximate the expression for the detection accuracy~\eqref{eq:generror1} with some convex function. This would reduce the complexity of numerical optimization, and may even allow for analytical results.

The effect of overfitting for medium back-off rates can be mitigated by choosing different back-off rates of the training stage and operational stage. For example, choosing larger back-off rates during training should increase the number of samples for states with many active nodes, thus reducing the risk of overfitting. Although this would simultaneously reduce the number of samples for smaller states, the risk of overfitting is not as high there due to the smaller number of active nodes.

Another direction for future research is to model temporal correlation in the sensor measurements in addition to spatial correlation.  In the present work, successive measurements in time are assumed independent, but including temporal correlation is more realistic \cite{VuranAA2004}.  If temporal correlation is part of the sensing and classification model, its interaction with the temporal back-off mechanism may produce quite interesting phenomena.  A Markov model for temporal correlation could be analyzed together with the Markov activity process of the CSMA model.

Finally, we also mention that asymptotic analysis is of interest in the future study of this cross-layer supervised learning and random-access communication setup.  Developing expressions for the three-node dependent network, e.g., \eqref{eqn:threenodem} and \eqref{eqn:threenodeA}, requires us to keep track of many details; larger networks will require us to keep track of many more.  By performing an asymptotic analysis of an increasing number of randomly placed sensor nodes with constant density in \cite{Varshney2012}, we are able to eliminate many such details in the sensor network generalization error using geometric probability \cite{PenroseY2003}.  Having now set forth this extended model with CSMA communication, similar asymptotic analysis using geometric probability is certainly warranted.



\IEEEtriggeratref{30}



%




\begin{thebibliography}{10}
\providecommand{\url}[1]{#1}
\csname url@samestyle\endcsname
\providecommand{\newblock}{\relax}
\providecommand{\bibinfo}[2]{#2}
\providecommand{\BIBentrySTDinterwordspacing}{\spaceskip=0pt\relax}
\providecommand{\BIBentryALTinterwordstretchfactor}{4}
\providecommand{\BIBentryALTinterwordspacing}{\spaceskip=\fontdimen2\font plus
\BIBentryALTinterwordstretchfactor\fontdimen3\font minus
  \fontdimen4\font\relax}
\providecommand{\BIBforeignlanguage}[2]{{%
\expandafter\ifx\csname l@#1\endcsname\relax
\typeout{** WARNING: IEEEtran.bst: No hyphenation pattern has been}%
\typeout{** loaded for the language `#1'. Using the pattern for}%
\typeout{** the default language instead.}%
\else
\language=\csname l@#1\endcsname
\fi
#2}}
\providecommand{\BIBdecl}{\relax}
\BIBdecl

\bibitem{NguyenWJ2005}
X.~Nguyen, M.~J. Wainwright, and M.~I. Jordan, ``Nonparametric decentralized
  detection using kernel methods,'' \emph{{IEEE} Trans. Signal Process.},
  vol.~53, no.~11, pp. 4053--4066, Nov. 2005.

\bibitem{PreddKP2006}
J.~B. Predd, S.~R. Kulkarni, and H.~V. Poor, ``Distributed learning in wireless
  sensor networks,'' \emph{{IEEE} Signal Process. Mag.}, vol.~23, no.~4, pp.
  56--69, Jul. 2006.

\bibitem{PreddKP2006a}
------, ``Consistency in models for distributed learning under communication
  constraints,'' \emph{{IEEE} Trans. Inf. Theory}, vol.~52, no.~1, pp. 52--63,
  Jan. 2006.

\bibitem{ZhengKP2011}
H.~Zheng, S.~R. Kulkarni, and H.~V. Poor, ``Attribute-distributed learning:
  Models, limits, and algorithms,'' \emph{{IEEE} Trans. Signal Process.},
  vol.~59, no.~1, pp. 386--398, Jan. 2011.

\bibitem{VarshneyW2011}
K.~R. Varshney and A.~S. Willsky, ``Linear dimensionality reduction for
  margin-based classification: High-dimensional data and sensor networks,''
  \emph{{IEEE} Trans. Signal Process.}, vol.~59, no.~6, pp. 2496--2512, Jun.
  2011.

\bibitem{Varshney2012}
K.~R. Varshney, ``Generalization error of linear discriminant analysis in
  spatially-correlated sensor networks,'' \emph{{IEEE} Trans. Signal Process.},
  vol.~60, no.~6, pp. 3295--3301, Jun. 2012.

\bibitem{AkyildizSSC2002}
I.~F. Akyildiz, W.~Su, Y.~Sankarasubramaniam, and E.~Cayirci, ``A survey on
  sensor networks,'' \emph{{IEEE} Commun. Mag.}, vol.~40, no.~8, pp. 102--114,
  Aug. 2002.

\bibitem{TayJB2004}
Y.~C. Tay, K.~Jamieson, and H.~Balakrishnan, ``Collision-minimizing {CSMA} and
  its applications to wireless sensor networks,'' \emph{{IEEE} J. Sel. Areas
  Commun.}, vol.~22, no.~6, pp. 1048--1057, Aug. 2004.

\bibitem{HongV2007}
Y.-W. Hong and P.~K. Varshney, ``Data-centric and cooperative {MAC} protocols
  for sensor networks,'' in \emph{Wireless Sensor Networks: Signal Processing
  and Communications Perspectives}, A.~Swami, Q.~Zhao, Y.-W. Hong, and L.~Tong,
  Eds.\hskip 1em plus 0.5em minus 0.4em\relax Chichester, UK: John Wiley \&
  Sons, 2007, pp. 311--348.

\bibitem{SinghB2012}
H.~Singh and B.~Biswas, ``Comparison of {CSMA} based {MAC} protocols of
  wireless sensor networks,'' \emph{Int. J. AdHoc Netw. Syst.}, vol.~2, no.~2,
  pp. 11--20, Apr. 2012.

\bibitem{802154}
802.15.4, ``{IEEE Standard for Local and metropolitan area networks--Part 15.4:
  Low-Rate Wireless Personal Area Networks},'' 2006.

\bibitem{RaudysY2004}
{\v{S}}.~Raudys and D.~M. Young, ``Results in statistical discriminant
  analysis: A review of the former {S}oviet {U}nion literature,'' \emph{J.
  Multivariate Anal.}, vol.~89, no.~1, pp. 1--35, Apr. 2004.

\bibitem{Vapnik1995}
V.~N. Vapnik, \emph{The Nature of Statistical Learning Theory}.\hskip 1em plus
  0.5em minus 0.4em\relax New York, NY: Springer, 1995.

\bibitem{FeoD2011}
E.~Feo and G.~A. Di~Caro, ``An analytical model for {IEEE} 802.15.4 non-beacon
  enabled {CSMA/CA} in multihop wireless sensor networks,'' Istituto Dalle
  Molle di Studi sull'Intelligenza Artificiale, Lugano, Switzerland, Tech. Rep.
  05-11, May 2011.

\bibitem{VuranA2006}
M.~C. Vuran and I.~F. Akyildiz, ``Spatial correlation-based collaborative
  medium access control in wireless sensor networks,'' \emph{{IEEE/ACM} Trans.
  Netw.}, vol.~14, no.~2, pp. 316--329, Apr. 2006.

\bibitem{ChamberlandV2003}
J.-F. Chamberland and V.~V. Veeravalli, ``Decentralized detection in sensor
  networks,'' \emph{{IEEE} Trans. Signal Process.}, vol.~51, no.~2, pp.
  407--416, Feb. 2003.

\bibitem{VuranAA2004}
M.~C. Vuran, {\"{O}}.~B. Akan, and I.~F. Akyildiz, ``Spatio-temporal
  correlation: Theory and applications for wireless sensor networks,''
  \emph{Comput. Netw.}, vol.~45, no.~3, pp. 245--259, Jun. 2004.

\bibitem{JindalP2006}
A.~Jindal and K.~Psounis, ``Modeling spatially correlated data in sensor
  networks,'' \emph{ACM Trans. Sensor Netw.}, vol.~2, no.~4, pp. 466--499, Nov.
  2006.

\bibitem{BK80}
R.~Boorstyn and A.~Kershenbaum, ``Throughput analysis of multihop packet
  radio,'' in \emph{Proc. Int. Conf. Commun.}, Seattle, WA, Jun. 1980, pp.
  1361--1366.

\bibitem{BKMS87}
R.~Boorstyn, A.~Kershenbaum, B.~Maglaris, and V.~Sahin, ``Throughput analysis
  in multihop {CSMA} packet radio networks,'' \emph{{IEEE} Trans. Commun.},
  vol. COM-35, no.~3, pp. 267--274, Mar. 1987.

\bibitem{WK05}
X.~Wang and K.~Kar, ``Throughput modelling and fairness issues in {CSMA/CA}
  based ad-hoc networks,'' in \emph{Proc.\ INFOCOM}, Miami, FL, Mar. 2005, pp.
  23--34.

\bibitem{LKLW08}
S.~Liew, C.~Kai, J.~Leung, and B.~Wong, ``Back-of-the-envelope computation of
  throughput distributions in {CSMA} wireless networks,'' \emph{{IEEE} Trans.
  Mobile Comput.}, vol.~9, no.~9, pp. 1319--1331, Sep. 2010.

\bibitem{DDT09}
M.~Durvy, O.~Dousse, and P.~Thiran, ``Self-organization properties of {CSMA/CA}
  systems and their consequences on fairness,'' \emph{{IEEE} Trans. Inf.
  Theory}, vol.~55, no.~3, pp. 931--943, Mar. 2009.

\bibitem{VandevenLDJ2012}
P.~M. van~de Ven, J.~S.~H. van Leeuwaarden, D.~Denteneer, and A.~J. E.~M.
  Janssen, ``Spatial fairness in linear random-access networks,''
  \emph{Perform. Evaluation}, vol.~69, no. 3--4, pp. 121--134, Mar.--Apr. 2012.

\bibitem{RSS09a}
S.~Rajagopalan, D.~Shah, and J.~Shin, ``Network adiabatic theorem: an efficient
  randomized protocol for content resolution,'' in \emph{Proc.\ ACM
  SIGMETRICS/Performance}, Seattle, WA, Jun. 2009, pp. 133--144.

\bibitem{BF10}
T.~Bonald and M.~Feuillet, ``On the stability of flow-aware {CSMA},''
  \emph{Perform. Evaluation}, vol.~67, no.~11, pp. 1219--1229, Nov. 2010.

\bibitem{GS10}
J.~Ghaderi and R.~Srikant, ``On the design of efficient {CSMA} algorithms for
  wireless networks,'' in \emph{Proc. IEEE Conf. Decision Control}, Atlanta,
  GA, Dec. 2010, pp. 954--959.

\bibitem{JW10a}
L.~Jiang and J.~Walrand, ``A distributed {CSMA} algorithm for throughput and
  utility maximization in wireless networks,'' \emph{{IEEE/ACM} Trans. Netw.},
  vol.~18, no.~3, pp. 960--972, Jun. 2010.

\bibitem{Bianchi00}
G.~Bianchi, ``Performance analysis of the {IEEE} 802.11 distributed
  coordination function,'' \emph{{IEEE} J. Sel. Areas Commun.}, vol.~18, no.~3,
  pp. 535--547, Mar. 2000.

\bibitem{PKCCK05}
T.~Park, T.~Kim, J.~Choi, S.~Choi, and W.~Kwon, ``Throughput and energy
  consumption analysis of {IEEE} 802.15.4 slotted {CSMA/CA},'' \emph{Electron.
  Lett.}, vol.~41, no.~18, pp. 1017--1019, Sep. 2005.

\bibitem{Fisher1936}
R.~A. Fisher, ``The use of multiple measurements in taxonomic problems,''
  \emph{Ann. Eugenics}, vol.~7, pp. 179--188, 1936.

\bibitem{Anderson1958}
T.~W. Anderson, \emph{Introduction to Multivariate Statistical Analysis}.\hskip
  1em plus 0.5em minus 0.4em\relax New York, NY: John Wiley \& Sons, 1958.

\bibitem{AnandkumarTS2009}
A.~Anandkumar, L.~Tong, and A.~Swami, ``Detection of {G}auss--{M}arkov random
  fields with nearest-neighbor dependency,'' \emph{{IEEE} Trans. Inf. Theory},
  vol.~55, no.~2, pp. 816--827, Feb. 2009.

\bibitem{GelmanH2007}
A.~Gelman and J.~Hill, \emph{Data Analysis Using Regression and
  Multilevel/Hierarchical Models}.\hskip 1em plus 0.5em minus 0.4em\relax
  Cambridge, UK: Cambridge University Press, 2007.

\bibitem{PenroseY2003}L
M.~D. Penrose and J.~E. Yukich, ``Weak laws of large numbers in geometric
  probability,'' \emph{Ann. Appl. Prob.}, vol.~13, no.~1, pp. 277--303, Jan.
  2003.

\end{thebibliography}
\end{document}